\documentclass[12pt,preprint]{aastex}

\shorttitle{\sc disk and excitation effects around HW~2}
\shortauthors{\sc Torrelles et al.}

\begin{document}


\title{The Circumstellar Structure and Excitation Effects around the Massive Protostar Cepheus A HW~2}


\author{J. M. Torrelles\altaffilmark{1}, N. A. Patel\altaffilmark{2}, S. Curiel\altaffilmark{3},
P. T. P. Ho\altaffilmark{2,4}, G. Garay\altaffilmark{5}, L. F. Rodr\'{\i}guez\altaffilmark{6}}

\altaffiltext{1}{Instituto de Ciencias del Espacio (CSIC)-IEEC, Facultat de F\'{\i}sica, 
Universitat de Barcelona, Mart\'{\i} i Franqu\`es 1, 08028 Barcelona, Spain; torrelles@ieec.fcr.es}
\altaffiltext{2}{Harvard-Smithsonian Center for Astrophysics, 60 Garden Street, Cambridge, MA 02138; npatel@cfa.harvard.edu, pho@cfa.harvard.edu}
\altaffiltext{3}{Instituto de Astronom{\'\i}a (UNAM), Ap. P. 70-264, M\'exico D.F., Mexico; scuriel@astroscu.unam.mx}
\altaffiltext{4}{Academia Sinica Institute of Astronomy and Astrophysics, Taipei, Taiwan}
\altaffiltext{5}{Departamento de Astronom\'{\i}a, Universidad de Chile, Casilla 36-D, Santiago, Chile; guido@das.uchile.cl}
\altaffiltext{6}{Centro de Radioastronom\'{\i}a y Astrof\'{\i}sica (UNAM), Morelia 58089, Mexico; l.rodriguez@astrosmo.unam.mx}

\begin{abstract}
  
We report SMA 335 GHz continuum observations with angular resolution
of $\sim$ $0\rlap.{''}3$, together with VLA ammonia
observations with $\sim$ $1''$ resolution toward
Cep A HW~2. We find that the flattened disk structure of the dust emission observed by Patel et al. is preserved at the $0\rlap.{''}3$ scale, showing an elongated
structure of $\sim$ $0\rlap.{''}6$ size (450 AU) peaking
on HW~2.  In addition, two ammonia cores are observed, one associated with a hot-core previously reported, and an elongated core with a double peak separated by $\sim$ $1\rlap.{''}3$
and with signs of heating  at the inner edges of the gas
facing HW~2. The double-peaked ammonia structure, as well as the
double-peaked CH$_3$CN structure reported previously (and proposed
to be two independent hot-cores), surround both the dust emission as well as the
double-peaked SO$_2$ disk structure
found by Jim\'enez-Serra et al. All these results argue against
the interpretation of the elongated  dust-gas structure as
due to a chance-superposition of different cores; instead,
they imply that it
is physically related to the central massive object within a {\it disk-protostar-jet} system.

\end{abstract}


\section{Introduction}

HW~2 is the brightest of the radio continuum objects detected in
the star forming region of Cepheus A (725 pc distance, Johnson 1957)
harboring a massive young star (probably a B0.5 zero-age main
sequence star; Hughes \& Wouterloot 1984, Garay et al. 1996). This
object, in addition to the strong radio continuum emission seen at
$\lambda$cm, is also related to well-known signatures typical of
young massive stars, such as bright masers, and intense magnetic fields
(e.g., Torrelles et al. 1996, 2001, Minier, Booth, \&
Conway 2000, Gallimore et al. 2003, Niezurawska et al. 2004,
Bartkiewicz et al. 2005, Vlemmings et al. 2006, Patel et al. 2007).
However, what makes this a unique object is its association with
similar phenomena as has been observed toward low-mass young stellar
objects. The HW~2 radio continuum jet detected at
the base ($\sim 1''$) appears to be driving the more extended
($\sim$ 1$'$) bipolar molecular outflow seen in HCO$^+$ (Rodr\'{\i}guez
et al. 1994, G\'omez et al. 1999). Especially remarkable are the large
proper motions observed in the two main components of the HW~2 radio
jet, moving away at $\sim$ 500~km~s$^{-1}$ from the central source
in nearly opposite directions and parallel to the HCO$^+$ bipolar
outflow (Curiel et al. 2006). These observations strongly
support theoretical models of high-mass star formation through an
accretion disk (McKee \& Tan 2003, De Buizer, Osorio, \& Calvet
2005, Beltr\'an et al. 2006, Banerjee \& Pudritz 2007), with the
ejection of bipolar outflows sharing similar characteristics with
the star-formation process of low-mass stars (Shu, Adams, \& Lizano
1987). This is in contrast to models requiring merging of low-mass stars
(Bonnell \& Bate 2005) where collimated jets are not expected.

The detection of a rotating disk of dust and molecular gas of $\simeq$ 330 AU radius oriented
perpendicular to, and spatially coincident with, the HW~2 radio
jet, has been reported through Submillimeter Array (SMA)
observations with an angular resolution of $\simeq$
$0\rlap.{''}75$ (Patel et al. 2005). Those results gave additional support to the picture that during the formation
of this massive object a {\it disk-protostar-jet} system has been formed.  Moreover, Jim\'enez-Serra et al. (2007),
through a detailed analysis of Very Large Array (VLA) and
Plateau de Bure Interferometer (PdBI) subarcsecond ($\sim$
$0\rlap.{''}2$-$0\rlap.{''}6$) $\lambda$mm observations, have resolved for the
first time the hot gas surrounding HW~2. 
They find that the radiation from HW~2 could be photoevaporating the disk and that the disk does
not appear to be rotating  with a Keplerian law due to the extreme youth of the object.
The size and kinematics of the SO$_2$ disk
inferred by Jim\'enez-Serra et. al. is consistent with the conclusions of Patel
et al. However, an alternative interpretation is that
the disk structure and kinematics observed are due
to the superposition on the plane of the sky of 
independent 
hot-cores (Comito et al. 2007, Brogan et al. 2007). Here we report new SMA
continuum observations and VLA NH$_3$(3,3) and NH$_3$(4,4)
observations, both with very high angular resolution, giving 
further support to the circumstellar disk interpretation.

\section{Observations and results}
\subsection{SMA continuum}
The submillimeter continuum observations were carried out in two
configurations of the SMA (Ho et al. 2004)\footnote{The Submillimeter Array is a
joint project between the Smithsonian Astrophysical Observatory
and the Academia Sinica Institute of Astronomy and Astrophysics,
and is funded by the Smithsonian Institution and the Academia
Sinica.}, the extended configuration with a maximum  baseline length
of 220 m, on 30 August 2004, and the very extended configuration with a  maximum
baseline length of 472 m, on 4 November 2005.  We summarize here
the observations of this latter date (see Patel et  al. 2005, 2007
for the details on the observations in the extended  configuration).
We used a tuning of 331.1 GHz centered in the lower sideband (and 341.1 GHz centered in the upper sideband). The quasar BL Lac was observed for 5 minutes
between every cycle of 10 minutes on the main source.  The spectral
band-pass was calibrated using observations of Mars and absolute
flux calibration was done using the asteroid Ceres and the quasar
3C111.  Continuum data were obtained from the line-free regions of
the  spectral band of the two sidebands.  Weather was excellent
during the observations with relative humidity of $\sim$ 20\% and
$\tau_{225 GHz}\sim 0.06$ ($\sim$ 0.22 @ 335~GHz), measured at the nearby Caltech Submillimeter
Observatory. The track was $\sim$~8 h long with an on-source integration
time of $\sim$~4.2 h. $T_{sys,DSB}$ varied from 220 to 500 K. The
visibility data were calibrated  and mapped   using the Berkeley
Illinois Maryland Array's Miriad package. We estimate an uncertainty
of $\sim$~20\% in the absolute flux scale in the SMA data and an uncertainty of $\sim$ $0\rlap.{''}1$ in absolute astrometry. Final maps were obtained by combining the two configurations (extended + very extended) and the two sidebands. Two continuum sources are detected with a beam size of $0\rlap.{''}37$, one coincident with HW~2 (S$_{(335~GHz)}$ $\simeq$ 0.6~Jy~beam$^{-1}$) and the another one with HW~3c (S$_{(335~GHz)}$ $\simeq$ 0.1~Jy~beam$^{-1}$), which is located $\sim$ 3$''$ south from HW~2. The continuum emission around HW~2 shows (Fig. 1) an elongated structure with a deconvolved size of $\simeq$ $0\rlap.{''}62\times0\rlap.{''}35$
(450$\times$250~AU; PA $\simeq$ 120$^{\circ}$) and 
total flux density of $\simeq$ 2~Jy (brightness temperature T$_B$ $\simeq$ 100~K) (similar to the flux density measured with a beam size of $\simeq$ $0\rlap.{''}75$ by Patel et al. 2005). A weak feature (6$\sigma$ level, $\sigma$ $\simeq$ 7~mJy/beam) located $\sim$ 1$''$ north from the peak of the elongated structure is also observed.
The implications of these high-angular resolution continuum observations are discussed in \S~3.
An analysis of the SMA spectral line data combining the two configurations
will be presented in a forthcoming paper by Patel et al.
\subsection{VLA ammonia}

We have reanalyzed the NH$_3$(3,3) and NH$_3$(4,4) line data
($\lambda$1.3cm) observed by Torrelles et al. (1999) who reported
detection of the lower transition but non-detection of the higher
one. These observations were carried out with the VLA of the
NRAO\footnote{The National Radio Astronomy Observatory is a facility
of the National Science Foundation operated under cooperative
agreement by Associated Universities, Inc.} in the C configuration
and 4IF spectral line mode, allowing us to observe simultaneously the
two transitions with a spectral resolution of $\simeq$ 0.6~km~s$^{-1}$
and covering a velocity range of --30 $\leq$ V$_{LSR}$ $\leq$
8~km~s$^{-1}$ (covering only the main hyperfine component of these transitions; see Ho \& Townes 1983). 
More details of these observations are given in
Torrelles et al. (1999).  The re-calibration of the observations
was performed using the new recommended procedure of NRAO for
reducing high-frequency VLA data. The continuum emission contribution from 
HW~2 to the spectral line data was subtracted.  Emission from both transitions 
is detected in the velocity range from $\simeq$ --13.6 to --2.6~km~s$^{-1}$ 
with a beam size of $\sim$ 1$''$.
The integrated flux density images in this range are
consistent with those reported by Torrelles et al. (1999) [NH$_3$(3,3)] 
and Brogan et al. (2007) [NH$_3$(4,4)], showing a rather complex structure 
around HW~2.
However, more simple structures are differentiated from integrated
intensity maps made in two different velocity ranges, from --13.6 to
--8.1~km~s$^{-1}$  and from --7.5 to --2.6~km~s$^{-1}$. 
In the former range of integration, a distinct spatially ``isolated'' 
ammonia core is identified in both transitions, spatially coinciding
with the hot-core reported by Mart\'{\i}n-Pintado et al. (2005) in SO$_2$ 
$\sim$ $0\rlap.{''}5$ to the east of HW~2 (Fig. 2). This core is also seen in CH$_3$CN (Comito et al. 2007).
On the other hand, an elongated core centered on HW~2 is
distinguished in the
--7.5 to --2.6~km~s$^{-1}$ range 
(more prominent in the NH$_3$(4,4) transition; Fig. 2). 
This structure has a similar
orientation as the CH$_3$CN and SO$_2$ structures reported by
Patel et al. (2005) and Jim\'enez-Serra et al. (2007), respectively,  
within the same velocity range, although the ammonia structure is 
about two times larger ($\sim$ 2$''$). In addition, the ammonia emission in the 
--7.5 to --2.6~km~s$^{-1}$ range shows a velocity pattern which is roughly consistent 
with the pattern observed in SO$_2$ by Jimenez-Serra et al. (2007); that is, the gas of the northwestern part of the elongated structure appears to be redshifted with respect to the southeastern part.
However, the S/N ratio of the individual ammonia velocity 
channels is not good enough (S/N $\lesssim$ 5-6~$\sigma$) to make a detailed kinematical
comparison, even more considering the different angular resolution of both 
observations (NH$_3$ and SO$_2$). 

An additional relevant result is obtained when a spatial comparison
of the (4,4) and (3,3) ammonia emission is made, given that their
flux density ratio is sensitive to the rotational temperature
describing the population of these two rotational states (Ho \&
Townes 1983).  In fact, from Fig. 2 we see that in the --13.6 to
--8.1~km~s$^{-1}$ range the (4,4) and (3,3) peaks of the ammonia
clumps do not coincide, but are separated by  $0\rlap.{''}2$, with
the (4,4) peak being closer to HW~2, suggesting external heating effects
by the massive protostar. A relative spatial displacement between
the (4,4) and (3,3) emission is also  observed in the
elongated structure (--7.5 to --2.6~km~s$^{-1}$ range),
with the two (4,4) peaks separated by $1\rlap.{''}2$ and being closer to HW~2 than the corresponding (3,3)
ammonia peaks which are separated by $1\rlap.{''}4$, suggesting that the 
inner edges of the
elongated structure facing HW~2 is being heated by the massive protostar. 
This relative spatial displacement is significant
considering that the accuracy in the relative position of the (4,4) and (3,3) ammonia emission is estimated to be $\sim$ $0\rlap.{''}05$. 
From the (4,4) to (3,3) ratios ($\gtrsim$ 0.4, obtained from the spectra),
assuming optically thin emission,
we estimate overall rotational temperatures T$_R$(4,4;3,3)
$\gtrsim$ 160~K for the elongated structure. These temperatures are consistent with those obtained by Mart\'{\i}n-Pintado et al.
(2005) through SO$_2$ observations and with the brightness dust temperature of $\sim$ 100~K (\S~2.1).

\section{Discussion}

Our new high-angular resolution SMA observations show that the 
flattened disk-like structure
of the dust emission observed previously by Patel et al. (2005) is still
preserved at $\sim$ $0\rlap.{''}37$ scale, with a single continuum
peak on HW~2. In fact, the dust continuum peak coincides in absolute position within  $0\rlap.{''}07$ with the HW~2 position
estimated at cm wavelengths by Curiel et al. (2006). In addition, this elongated structure is similar in size and orientation
to the molecular SO$_2$ disk structure detected by Jim\'enez-Serra et
al. (2007) (Fig. 1), but with the dust continuum peaking in between the two peaks of the SO$_2$ structure. Both the dust continuum
and SO$_2$ structures are engulfed by the elongated double-peaked ammonia structure as well as by the double-peaked
CH$_3$CN structure detected at 230 GHz with arcsecond resolution
(proposed to be delineating two hot cores,
HC2 and HC3; Comito et al. 2007) (Fig. 1). Our results argue against the interpretation that the elongated structure 
observed perpendicular to the
HW~2 jet and proposed to be a disk by Patel et al. (2005) is due to a
chance superposition of different hot-cores (HC~2 and HC~3) given
that they have an angular separation ($\sim$ $1"$)
that is significantly larger than the size of the
elongated dust continuum structure ($\sim$ $0\rlap.{''}6$). Furthermore, 
the fact that a main single continuum peak on HW~2 is observed,
gives additional evidence that the observed dust continuum emission
distribution is dominated by the gas directly associated with HW~2.
We note that notwithstanding the importance of chemistry, with molecules being enhanced and depleted depending
on the physical conditions, dust emission is much less affected by such
chemical effects and hence is a more reliable indicator of the
structure of material in the immediate vicinity of HW~2.
We would like to emphasize, however, that we are not discarding the presence of multiple hot-cores
in the immediate vicinity of HW~2. Furthermore, these hot-cores seem to be necessary
to explain the observed chemical inhomogeneities in the region 
(Comito et al. 2007, Brogan et al. 2007). However, what we find here is that the SMA 
dust disk structure is not made up of these reported hot-cores.

The question arises as to the meaning of the double-peaked molecular
structure observed in SO$_2$ (Jim\'enez-Serra et al.) inside of
the double-peaked molecular structures observed in CH$_3$CN (Comito et al. 2007) and in 
NH$_3$ (this paper), all these structures observed within the same integrated velocity range ($\sim$ --7.5 to --2.6~km~s$^{-1}$; Fig. 1). As far as we know, this is the first time that this behaviour
has been observed and reported at (sub)arcsecond scale toward a high-mass star forming region.
As an open issue,
we discuss here three different
interpretations, not necessarily independent. The first one is that each of the different molecular
peaks are tracing molecular fragments from the remnant of the parental
core from which the central massive object has been formed.
The fact that all of them are almost aligned along the same axis,
roughly perpendicular to the HW~2 jet, would favor this possibility.
In addition, the observed heating of
the gas facing HW~2 (\S 2.2),
indicates a physical association of the
large (2$''$) elongated molecular structure with the massive
protostar, rather than a projection effect by the superposition in
the plane of the sky of different molecular cores.
A second possible explanation is that chemical and
excitation effects produce different spatial molecular peaks
for different molecular transitions.
This could be consistent with the fact that SO$_2$ (observed closer to HW ~2) trace
regions with densities two order of magnitud higher than the ammonia (van der Tak et al. 2007).
This would also be consistent with the fact first noted by Jim\'enez-Serra et al. (2007) that
the water masers reported by Torrelles et al. (1996) are distributed along the SO$_2$ emission
(water masers arise in regions with very high densities, $\sim$ 10$^9$~cm$^{-3}$; Elitzur 1989).
A third possible explanation is that a more 
or less continuous structure but with chemical variations becomes denser and thinner as we 
probe closer to the protostar, giving rise to concentric flattened 
structures of similar shape and orientation.
A beam-averaged radius of the disk growing with increasing beam size
would then produce the concentric double-peaked structures. 
We hypothesize that this could be the case for HW~2 given that 
the angular separation of the two peaks of the different molecular structures found around HW~2 
decreases with decreasing beam size (Fig. 3).
This possible scenario might be tested
by observing with the VLA the ammonia emission with higher angular resolution ($\sim$ $0\rlap.{''}3$). 
If this scenario is correct, a smaller separation between the corresponding double ammonia peaks should be measured. On the contrary, if the spatial separation
between the ammonia peaks does not change when observed with higher angular resolution, it would indicate that chemical and excitation effects are the dominant effects in producing the different double-peaked
molecular structures.

The main heating mechanism of the gas
around HW~2, 
can be explained
via collisions with hot dust, which in turn is heated by the absorption
of the radiation of this high-mass protostar (although a shock-heating
contribution from the HW~2 wind and/or from the external material infalling into the disk is also expected). From
Scoville \& Kwan (1976) and Zhang et al. (2007), we find that the expected dust temperature at a distance of $0\rlap.{''}3$
(220~AU) from HW~2 is T$_d$ $\simeq$ 100-180~K (depending on the grain emissivity spectral index $\beta$ = 2 to 1) for a source
with a luminosity of $\sim$ 2.5$\times$10$^4$~L$_{\odot}$ (Evans et
al. 1981, Lenzen et al. 1984).  Since collisional coupling of the
gas and dust is expected for densities n(H$_2$) $\gtrsim$
10$^5$~cm$^{-3}$ (Goldsmith \& Langer 1978), a value consistent
with the detection of NH$_3$, CH$_3$CN, and SO$_2$ emission, we
conclude that molecular gas heating up to the observed temperatures and distances
is possible. Comito et al. (2007) conclude that the mm/submm portion of the continuum emission is
optically thick. Therefore, to calculate the gas mass of the disk we follow the procedure outlined
by Beuther et al. (2007), assuming optically thin emission and correct that 
through the opacity correction factor C = $\tau$/(1-e$^{-\tau}$), with $\tau$ the dust opacity.
In this way we estimate the gas mass to be (0.4-3)$\times$C~M$_{\odot}$,
assuming a gas to dust ratio of 100, temperature of $\sim$ 160~K, and depending on the grain emissivity spectral index $\beta$ = 1 to 2.

\acknowledgments
We are grateful to the SMA staff in Cambridge, Hilo and Taipei, for
their help with observations.  
We thank I. Jim\'enez-Serra and C. Comito for kindly providing respectively the SO$_2$ and
CH$_3$CN images shown in Fig. 1. We also thanks J. M. Girart and P. F. Goldsmith for their very valuable comments.
The excellent comments and suggestions of an anonymous referee helped improve this Letter. SC acknowledges support from DGAPA/UNAM
and from CONACyT (M\'exico) grant  43120$-$F. G.G. acknowledges support from FONDAP 15010003 (Chile). JMT acknowledges support from the Spanish grant AYA2005-08523-C03.



\clearpage

\begin{figure}
\includegraphics[angle=-90,width=6.5in]{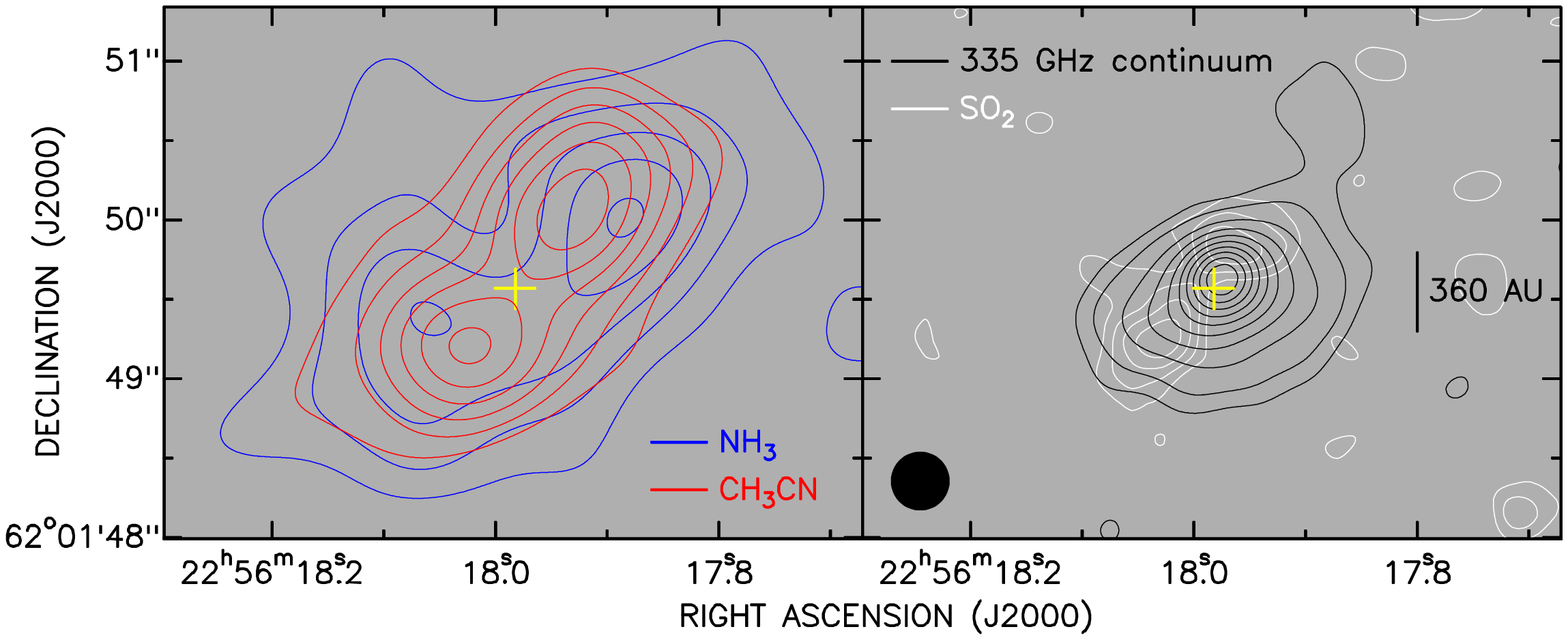}
\caption{{\it Left panel}: Contour map of the integrated ammonia emission of the (4,4) transition (blue contours) in the velocity range from --7.5 to --2.6~km~s$^{-1}$ (beam $\simeq$ $1\rlap.{''}1$, this paper; see Fig. 2) superposed on the 
CH$_3$CN contour map (red contours) obtained by Comito et al. (2007) within the same velocity range (beam $\simeq$ $0\rlap.{''}8$).
Cross indicates the position of HW~2 (Curiel et al. 2006).  {\it Right panel}: Contour map (black contours) of the continuum emission at 335~GHz obtained with the SMA toward Cep A HW~2 (this paper). Beam size = $0\rlap.{''}37$ is indicated in the panel.
Contour levels are --0.5, 0.5, 1, 2, 3, 4, 5, 6, 7, 8, 9 times 58~mJy/beam. Superposed on this map the SO$_2$ structure found
with the VLA by Jim\'enez-Serra et al. (2007) is also shown (white contours; 
beam $\simeq$ $0\rlap.{''}37$).} 
\end{figure}

\clearpage

\begin{figure}
\includegraphics[angle=0, width=6.5in]{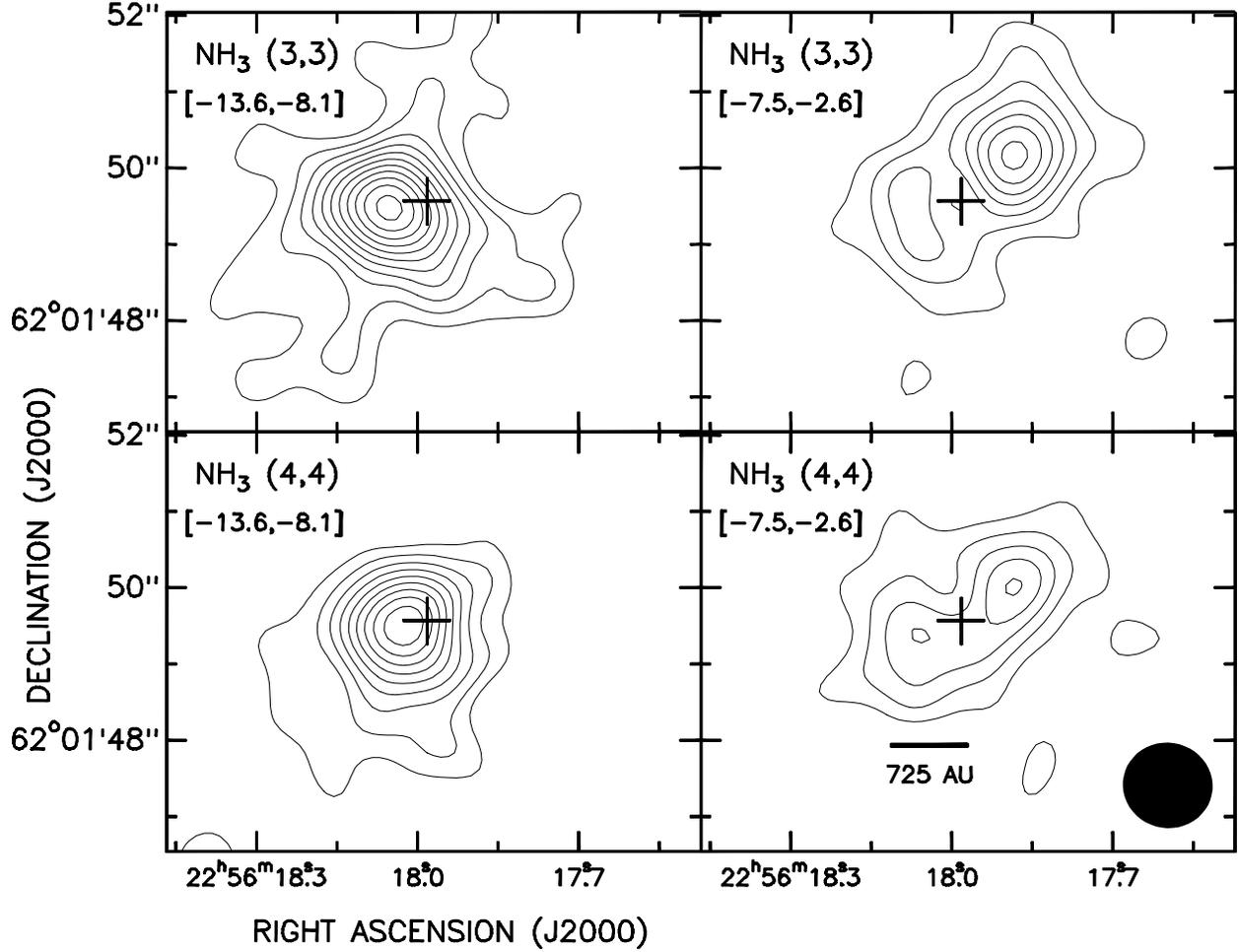}
\caption{Contour maps of the integrated flux density ammonia emission of the (3,3) and (4,4) inversion transition lines for different velocity ranges in km~s$^{-1}$ (indicated in the figures). Contour levels are 2, 3, 4, 5, 6, 7, 8, 9, 10, 11, and 12 times 14~mJy~beam$^{-1}$~km~s$^{-1}$.  
Beam size $\simeq$ $1\rlap.{''}1$ is indicated in the lower right panel.
The cross indicates the position of HW~2 (Curiel et al. 2006). The accuracy in the relative position of the (4,4) and (3,3) ammonia emission is estimated to be $\sim$ $0\rlap.{''}05$.} 
\end{figure}



\clearpage

\begin{figure}
\includegraphics[angle=-90,width=6.0in]{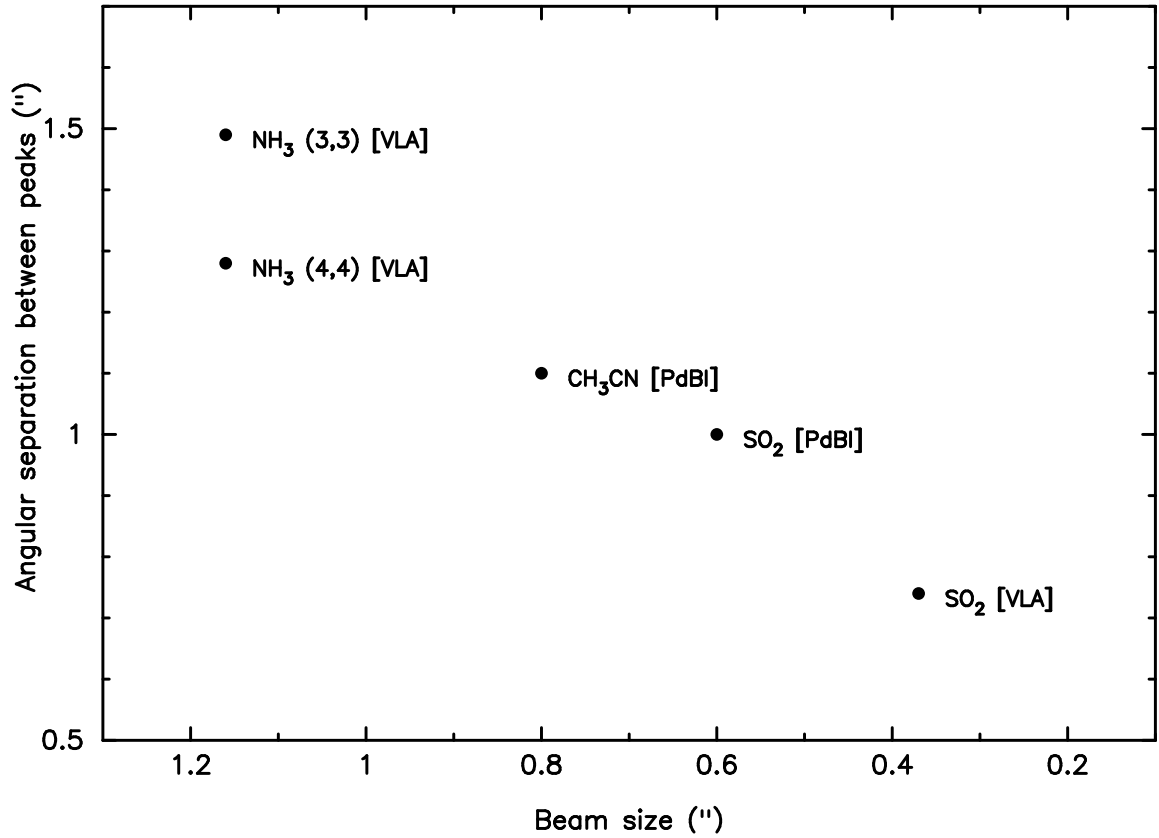}
\caption{Angular separation of the two peaks of the different molecular structures found around HW~2 as a function of the observing beam size. The separation decreases with decreasing beam size.
SO$_2$ (PdBI and VLA; Jim\'enez-Serra et al. 2007), CH$_3$CN (PdBI; Comito et al. 2007), NH$_3$ (VLA; this paper).}
\end{figure}

\end{document}